\documentclass[12pt]{article}

\usepackage{latexsym}
\usepackage{amssymb}
\usepackage{amsmath}

\def\Spin{{\rm Spin}}

\def\Pin{{\rm Pin}}

\def\SO{{\rm SO}}

\def\U{{\rm U}}

\def\su{{\rm su}}
\def\Mat{{\rm Mat}}
\def\C{{\mathbb C}}

\def\diag{{\rm diag}}

\def\R{{\mathbb R}}

\begin{document}

\title{A new nonlinear generalization of the Dirac equation}

\author{Nikolay Marchuk\footnote{Steklov Mathematical Institute, email: nmarchuk@mi.ras.ru}}

\maketitle

\begin{abstract}
We postulate a new nonlinear generalization of the Dirac equation for an electron. Basic properties of the new equation are considered.
\end{abstract}
\bigskip

MSC classes:	15A66, 15A75, 53Z05
\bigskip

\thanks{This work was partially supported by the grant of the President
of the Russian Federation (project NSh-2928.2012.1) and by Division
of mathematics of RAS (project ``Modern problems in theoretical
mathematics'').}
\bigskip

%%%%%%%%%%%%%%%%%%%%%%%%%%%%%%%%%%%%%%%%%%%%%%%%%%%%%%%%%%%%%%%%%%%%%%
%%%%%%%%%%%%%%%%%%%%%%%%%%%%%%%%%%%%%%%%%%%%%%%%%%%%%%%%%%%%%%%%%%%%%%%%%%%%%%%%%%%

%%%%%%%%%%%%%%%%%%%%%%%%%%%%%%%%%%%%%%%%%%%%%%%%%%%%%%%%%%%%%%%%%%%

\noindent{\bf Dirac equation for an electron.}
Let $\R^{1,3}$ be the Minkowski space with cartesian coordinates $x^\mu$, $\mu=0,1,2,3$, with partial derivatives $\partial_\mu=\partial/\partial x^\mu$, and with a metric tensor given by the diagonal matrix $\eta=\diag(1,-1,-1,-1)$.
Consider the Dirac equation for an electron
\begin{equation}
i\gamma^\mu(\partial_\mu\psi-i a_\mu\psi)-m\psi=0,\label{Dirac:eq}
\end{equation}
where $\psi=\psi(x)$ is a Dirac spinor, $a_\mu$ are components of a covector potential of
electromagnetic field, and
\begin{eqnarray*}
\gamma^0&=&\begin{pmatrix}1 &0 &0 & 0\cr
                  0 &1 & 0&0 \cr
                  0 &0 &-1&0 \cr
                  0 &0 &0 &-1\end{pmatrix},\quad
\gamma^1=\begin{pmatrix}0 &0 &0 & 1\cr
                  0 &0 & 1&0 \cr
                  0 &-1 &0 &0 \cr
                  -1 &0 &0 &0\end{pmatrix},
\\
\gamma^2&=&\begin{pmatrix}0 &0 &0 & -i\cr
                  0 &0 & i&0 \cr
                  0 & i&0 &0 \cr
                  -i &0 &0 &0\end{pmatrix},\quad
\gamma^3=\begin{pmatrix}0 &0 & 1& 0\cr
                  0 &0 & 0&-1 \cr
                  -1 &0 &0 &0 \cr
                  0 &1&0 &0\end{pmatrix}
\nonumber
\end{eqnarray*}
are $\gamma$-matrices in the Dirac representation. We have
$$
\gamma^\mu\gamma^\nu=-\gamma^\nu\gamma^\mu,\quad \mu\neq \nu,\quad
(\gamma^0)^2={\bf1},\quad
(\gamma^1)^2=(\gamma^2)^2=(\gamma^3)^2=-{\bf1},
$$
where ${\bf1}$ is the four dimensional identity matrix. Note that $i\gamma^\mu\in\su(2,2)$ (see \cite{Mybook:eng}).

Let us recall basic properties of the Dirac equation.

\begin{enumerate}
\item The electric charge conservation law:
$$
\partial_\mu j^\mu=0,\quad\hbox{where}\quad
j^\mu=\bar\psi\gamma^\mu\psi=\psi^\dagger\gamma^0\gamma^\mu\psi,
$$
where $\dagger$ is the operation of Hermitian conjugation.
\item Lorentz invariance of the Dirac equation:
$$
x^\mu\to\acute x^\mu=p^\mu_\nu x^\nu,\quad
\psi\to\acute\psi=S\psi,
$$
where $P=\|p^\mu_\nu\|\in O(1,3)$, $S\in\Pin(1,3)$ and $S^{-1}\gamma^\mu
S=p^\mu_\nu\gamma^\nu$. \item Gauge invariance w.r.t. $\U(1)$ gauge
Lie group:
$$
\psi\to\hat\psi=e^{i\lambda}\psi,\quad a_\mu\to\hat
a_\mu=a_\mu+\partial_\mu\lambda,\quad \lambda : \R^{1,3}\to\R.
$$
\item Decomposition of the Klein-Gordon-Fock operator:
\begin{equation}
(i\gamma^\mu\partial_\mu-m)(i\gamma^\nu\partial_\nu+m)=-(\partial_\mu\partial^\mu+m^2).\label{KGF:decomp}
\end{equation}
\end{enumerate}

%%%%%%%%%%%%%%%%%%%%%%%%%%%%%%%%%%%%%%%%%%%%%%%%%%%%%%%%%%%%%%%%%%%%%%%%%

\noindent{\bf Generalized Dirac equation with nonlinearity.}
Let us take ${\rm I}=\gamma^0\gamma^1\gamma^2\gamma^3$. We see that ${\rm
I}^2=-{\bf1}$, ${\rm I}^\dagger=-{\rm I}$.
Denote a subalgebra of matrix algebra
$$
N = \{\alpha{\bf1}+\beta {\rm I}\in\Mat(4,\C) :
\alpha,\beta\in\R\} \simeq\C.
$$
$$
Z=\alpha{\bf1}+\beta {\rm I}\leftrightarrow z=\alpha+i\beta\in\C.
$$

Let $f=f(z)$ be a continuous function $f : \C\to\C$  and let $F=F(Z)$  be the
corresponding function $F : N\to N$ such that $F(Z)|_{{\bf1}\to1, {\rm I}\to
i}=f(z)$.

Let us postulate the following equation, which depend on the function $F:N\to N$:
\begin{equation}
i\gamma^\mu(\partial_\mu\psi-i a_\mu\psi)-F(Z)\psi=0,\label{nonlin:Dirac}
\end{equation}
where
$$
Z=(\bar\psi\psi){\rm1}-(\bar\psi{\rm I}\psi){\rm I}
$$
The first term in the equation (\ref{nonlin:Dirac}) is equal to the first term in the Dirac equation for an electron (\ref{Dirac:eq}). So, we say that the equation (\ref{nonlin:Dirac}) is {\em a generalized Dirac equation (with a nonlinearity)}.

Consider basic properties of the generalized Dirac equation  (\ref{nonlin:Dirac}).
\begin{enumerate}
\item The electric charge conservation law:
$$
\partial_\mu j^\mu=0,\quad\hbox{where}\quad
j^\mu=\bar\psi\gamma^\mu\psi=\psi^\dagger\gamma^0\gamma^\mu\psi.
$$
\item Lorentz invariance of the Dirac equation:
$$
x^\mu\to\acute x^\mu=p^\mu_\nu x^\nu,\quad
\psi\to\acute\psi=S\psi,
$$
where $P=\|p^\mu_\nu\|\in\SO_+(1,3)$, $S\in\Spin_+(1,3)$ and $S^{-1}\gamma^\mu
S=p^\mu_\nu\gamma^\nu$. \item Gauge invariance w.r.t. $\U(1)$ gauge
Lie group:
$$
\psi\to\hat\psi=e^{i\lambda}\psi,\quad a_\mu\to\hat
a_\mu=a_\mu+\partial_\mu\lambda,\quad \lambda : \R^{1,3}\to\R.
$$
\item Decomposition of the second order operator:
\begin{equation}
(i\gamma^\mu\xi_\mu-F(Z))(i\gamma^\nu\xi_\nu+\overline{F(Z)})=-(\xi_\mu\xi^\mu+|F(Z)|^2).\label{KGFnew:decomp}
\end{equation}
If $F(Z)=\sigma{\bf1}+\rho I$, where $\sigma,\rho$ is functions $\R^{1,3}\to\R$, then $\overline{F(Z)}= \sigma{\bf1}-\rho I$, $|F(Z)|^2=\sigma^2+\rho^2$, and $\xi^\mu$ are commutative symbols.
\end{enumerate}
We see two differences in the basic properties of equations (\ref{nonlin:Dirac}) and (\ref{Dirac:eq}).
\begin{itemize}
\item The equation (\ref{Dirac:eq}) is invariant under Lorentz transformations of coordinates from the Lie group $O(1,3)$, but the equation (\ref{nonlin:Dirac}) is invariant under Lorentz transformations of coordinates from the proper orthochronous Lorentz group $\SO_+(1,3)$.

\item The decomposition (\ref{KGF:decomp}) is, generally speaking, different to the decomposition  (\ref{KGFnew:decomp}).
\end{itemize}

Consider special cases of the generalized Dirac equation.

 If we take $F(Z)\equiv m{\bf1}$ in (\ref{nonlin:Dirac}), then we get the Dirac equation (\ref{Dirac:eq}). That means the equation (\ref{nonlin:Dirac}) is, in fact, a generalization of the equation   (\ref{Dirac:eq}).

Let us remind that the Dirac equation (\ref{Dirac:eq}) can be derived from the Lagrangian
$$
{\cal L}=\bar\psi i\gamma^\mu(\partial_\mu\psi-i
a_\mu\psi)-m(\bar\psi\psi).
$$

If we take $F(Z)\equiv Z=(\bar\psi\psi){\rm1}-(\bar\psi{\rm I}\psi){\rm
I}$, then we get the equation
$$
i\gamma^\mu(\partial_\mu\psi-i a_\mu\psi)-
((\bar\psi\psi){\rm1}-(\bar\psi{\rm I}\psi){\rm I})\psi=0,
$$
which can be derived from the Lagrangian
$$
{\cal L}=\bar\psi i\gamma^\mu(\partial_\mu\psi-i
a_\mu\psi)-\frac{1}{2}(\bar\psi\psi)^2+\frac{1}{2}(\bar\psi{\rm I}\psi)^2.
$$

\medskip
\noindent{\bf Heisenberg's nonlinear field
equation.} One can see similarity between the generalized Dirac equation and the Heisenberg nonlinear field
equation \cite{Hei}
\begin{equation}
i\gamma^\mu(\partial_\mu\psi-i a_\mu\psi)-
(\bar\psi\gamma_\mu\psi)\gamma^\mu\psi-(\bar\psi\gamma_\mu{\rm
I}\psi)\gamma^\mu{\rm I}\psi=0.\label{Heis:eq}
\end{equation}

Heisenberg in \cite{Hei} had made an attempt to create a unified field theory on the basis of his equation (\ref{Heis:eq}).
So, it will be interesting to develop such a theory on the basis of new generalized Dirac equation  (\ref{nonlin:Dirac}).


\begin{thebibliography}{99}
\bibitem{Hei} Heisenberg W., Introduction to the unified field
theory of elementary particles, Intersience publishers, 1966.
\bibitem{Mybook:eng} Marchuk N., Field theory equations, Amazon, 2012.
\end{thebibliography}
\end{document}